\documentclass[aps,pra,twocolumn,superscriptaddress,groupedaddress,showpacs,floatfix]{revtex4}
\usepackage[dvips]{graphicx}
\usepackage{tabularx}
\usepackage{bm}
\usepackage{color}

\begin{document}

%\preprint{}

\title{Two-color photoassociation spectroscopy of ytterbium atoms and the precise determinations of $s$-wave scattering lengths}


\author{Masaaki Kitagawa}
\author{Katsunari Enomoto}
\author{Kentaro Kasa}
\affiliation{Department of Physics,
Graduate School of Science,
Kyoto University, Kyoto 606-8502, Japan}
\author{Yoshiro Takahashi}
\affiliation{Department of Physics,
Graduate School of Science,
Kyoto University, Kyoto 606-8502, Japan}
\affiliation{CREST, Japan Science and Technology Agency,
Kawaguchi, Saitama 332-0012, Japan}
\author{Roman Ciury\l{}o}
\affiliation{Instytut Fizyki, Uniwersytet Miko\l{}aja
Kopernika, ul. Grudzi\c{a}dzka 5/7, 87--100 Toru\'n, Poland}
\author{Pascal Naidon}
\affiliation{Atomic Physics Division and $^5$Joint Quantum Institute, 
National Institute of Standards and Technology,
100 Bureau Drive, Stop 8423, Gaithersburg, Maryland 20899-8423, USA.}
\author{Paul S. Julienne$^{4,5}$}

%\author{Pascal Naidon}
%\affiliation{Atomic Physics Division, National Institute of Standards and Technology,
%100 Bureau Drive, Stop 8423, Gaithersburg, Maryland 20899-8423, USA.}
%\author{Paul S. Julienne}
%\affiliation{Atomic Physics Division, National Institute of Standards and Technology,
%100 Bureau Drive, Stop 8423, Gaithersburg, Maryland 20899-8423, USA.}
%\affiliation{Atomic Physics Division and Joint Quantum Institute, National Institute of Standards and Technology,
%100 Bureau Drive, Stop 8423, Gaithersburg, Maryland 20899-8423, USA.}

\date{\today}

\begin{abstract}
By performing high-resolution two-color photoassociation spectroscopy,
we have successfully determined the binding energies of several of the last bound states of
the homonuclear dimers of six different isotopes of ytterbium. These spectroscopic data are
in excellent agreement with theoretical calculations based on a simple model potential,
which very precisely predicts the $s$-wave scattering lengths of all 28 pairs of the seven
stable isotopes.  The $s$-wave scattering lengths for collision of two atoms of the same
isotopic species are $13.33(18)$ nm for $^{168}$Yb, $3.38(11)$ nm for $^{170}$Yb, $-0.15(19)$ nm
for $^{171}$Yb, $-31.7(3.4)$ nm for $^{172}$Yb, $10.55(11)$ nm for $^{173}$Yb, $5.55(8)$ nm
for $^{174}$Yb, and $-1.28(23)$ nm for $^{176}$Yb.  The coefficient of the lead term of
the long-range van der Waals potential of the Yb$_2$ molecule is $C_6=1932(30)$
atomic units $(E_h a_0^6  \approx 9.573 \times 10^{-26}$ J nm$^6$).
\end{abstract}


\maketitle

\section{INTRODUCTION}

\begin{figure}
\begin{center}
\includegraphics[width=8cm,clip]{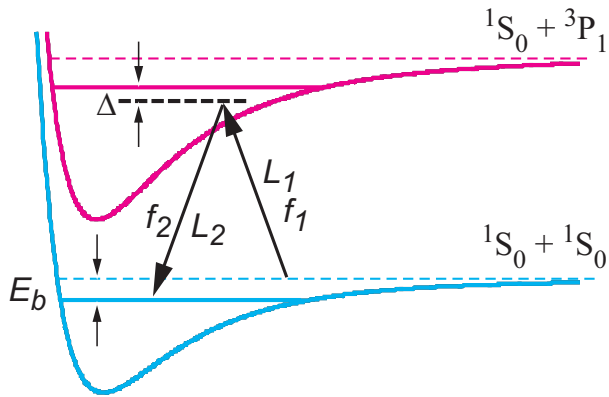}
\caption{(color online) Schematic description of the two-color PA spectroscopy. The laser $L_1$ drives one-color PA transition.
The laser $L_2$ couples the bound state in the excited molecular potential to the one in the ground molecular potential.
The detuning $\Delta$ of the PA laser with respect to the one-color PA resonance is set to several MHz for the Raman spectroscopy, while $\Delta$ is set to zero for the Autler-Townes spectroscopy.}
\label{fig:adiabatic potential}
\end{center}
\end{figure}

A collision between two atoms is a fundamental physical process and can be described by a few partial scattering waves for ultracold atoms.
At a sufficiently low temperature the kinetic energy of colliding atoms becomes less than the centrifugal barrier, and only the $s$-wave scattering is possible.
The $s$-wave scattering length is an essential parameter for describing ultracold collisions.
It also governs the static and dynamic properties of quantum degenerate gases like a Bose-Einstein condensate (BEC) or a degenerate Fermi gas (DFG) of fermionic atoms in different spin states.
Since the $s$-wave scattering length is very sensitive to the ground state interatomic potential,
especially at short internuclear distance, the precise {\it ab initio} calculation of the scattering length is very difficult, and therefore we must resort to experimental determination.
The most powerful approach for determining the scattering length is to measure the binding energy ($E_b$) of the last few bound states in the molecular ground state,
since the energy $E_b$ is closely related to the $s$-wave scattering length.
So far, the binding energies were measured via two-color photoassociation (PA) spectroscopy
for Li \cite{Li}, Na \cite{Na2}, K \cite{K}, Rb \cite{Rb}, Cs \cite{Cs} and He \cite{He1}.
A schematic description is shown in Fig.~\ref{fig:adiabatic potential}.
If a laser field $L_2$ is resonant to a bound-bound transition, it causes an Autler-Townes doublet.
This effect is detected as a reduction of a rate of a free-bound PA transition driven by the other laser field $L_1$.
This scheme is called Autler-Townes spectroscopy, and it is also explained in terms of the formation of a dark state.
If both lasers are off-resonant and the frequency difference matches $E_b$, these lasers drive a stimulated Raman transition from the colliding atom pair to a molecular state in the electronic ground state.
In this Raman spectroscopy, the resonance is detected as an atom loss.

The $s$-wave scattering length of two colliding atoms is determined by the adiabatic Born-Oppenheimer interaction potential $V(r)$ between the two atoms, 
which is very well approximated at large interatomic separation $r$ by the van der Waals contribution, $V(r)\approx -C_{6}/r^{6}$, where $C_6$ is the van der Waals  coefficient due to the dipole-dipole interaction.  The $s$-wave scattering length is given by the following formula, based on a quantum correction to the WKB approximation so as to be accurate for the zero-energy ($E=0$) limit \cite{GF1,Flambaum,Boisseau},
\begin{equation}
a = \bar{a} \left[ 1-\tan\left(\Phi-\frac{\pi}{8}\right)\right].
\label{massscaling}
\end{equation}
Here $\bar{a}=2^{-3/2}\frac{\Gamma(3/4)}{\Gamma(5/4)}\left(2\mu C_6/\hbar^2\right)^\frac{1}{4}$ is a characteristic length associated with the van der Waals potential,
where $\Gamma$ is the gamma-function, $\mu $ is the reduced mass, and $\hbar $ is the Planck constant divided by $2\pi $.
The semiclassical phase $\Phi$ is defined by
\begin{equation}
\Phi = \frac{\sqrt{2\mu}}{\hbar}\int_{r_0}^{\infty} \sqrt{-V(r)}dr,
\end{equation}
where $r_0$ is the inner classical turning point of $V(r)$ at zero energy.  The number of bound states $N$ in the potential is~\cite{Flambaum}
\begin{equation}
 N=\left [ \frac{\Phi}{\pi} -\frac{5}{8} \right ] +1,
 \label{Nbound}
\end{equation}
where $[ \ldots ]$ means the integer part.  As was mentioned above, the scattering length is very sensitive to the phase $\Phi $ and can take on any value between $\pm\infty$ when $\Phi$ varies over a range spanning $\pi$.  
While the accurate calculation of $\Phi $ is very difficult since it requires a knowledge of the whole potential, $\Phi$ is proportional to $\sqrt{\mu}$, 
and so the formula in Eq.~ (\ref{massscaling}) gives a simple mass-scaling of the isotopic variation of the scattering length once $a$ and $N$ are known for one isotopic combination. 
Mass scaling applies as long as small mass-dependent corrections to the Born-Oppenheimer potential~\cite{Bunker77,Wolniewicz83,Williams93} can be ignored. 
While mass scaling is often used for bound and scattering states for cold atomic systems ~\cite{vanKempen,Ferlaino}, and seems applicable to large-mass systems within experimental uncertainties ~\cite{Seto}, exceptions are known~\cite{Falke07}, and its accuracy should be carefully tested for different kinds of systems.

\begin{figure}
\begin{center}
\includegraphics[width=8cm,clip]{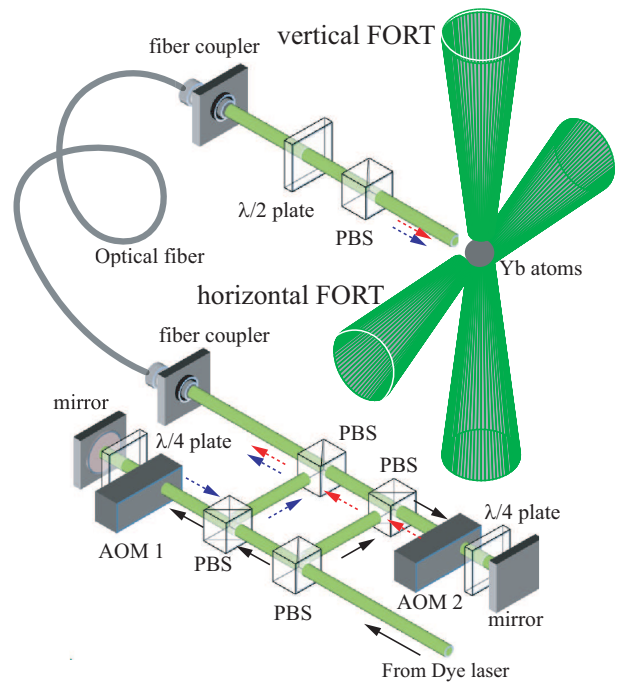}
\caption{(color online) Schematic description of an experimental setup for PA lasers.
The two lasers were prepared by splitting one laser, and
two double-pass acousto-optic modulators (AOMs) were employed to tune them.
Then the two lasers were aligned
in the same path.
The solid lines and broken lines indicate the laser beams before
and after the double-pass, respectively.
A polarizing beam splitter (PBS) and a $\lambda  /2$ plate after the optical fiber were employed to fix polarization of both lasers in the same direction.}
\label{fig:PAlaser}
\end{center}
\end{figure}
\begin{figure*}[t]
\begin{center}
\includegraphics[width=1.0\hsize]{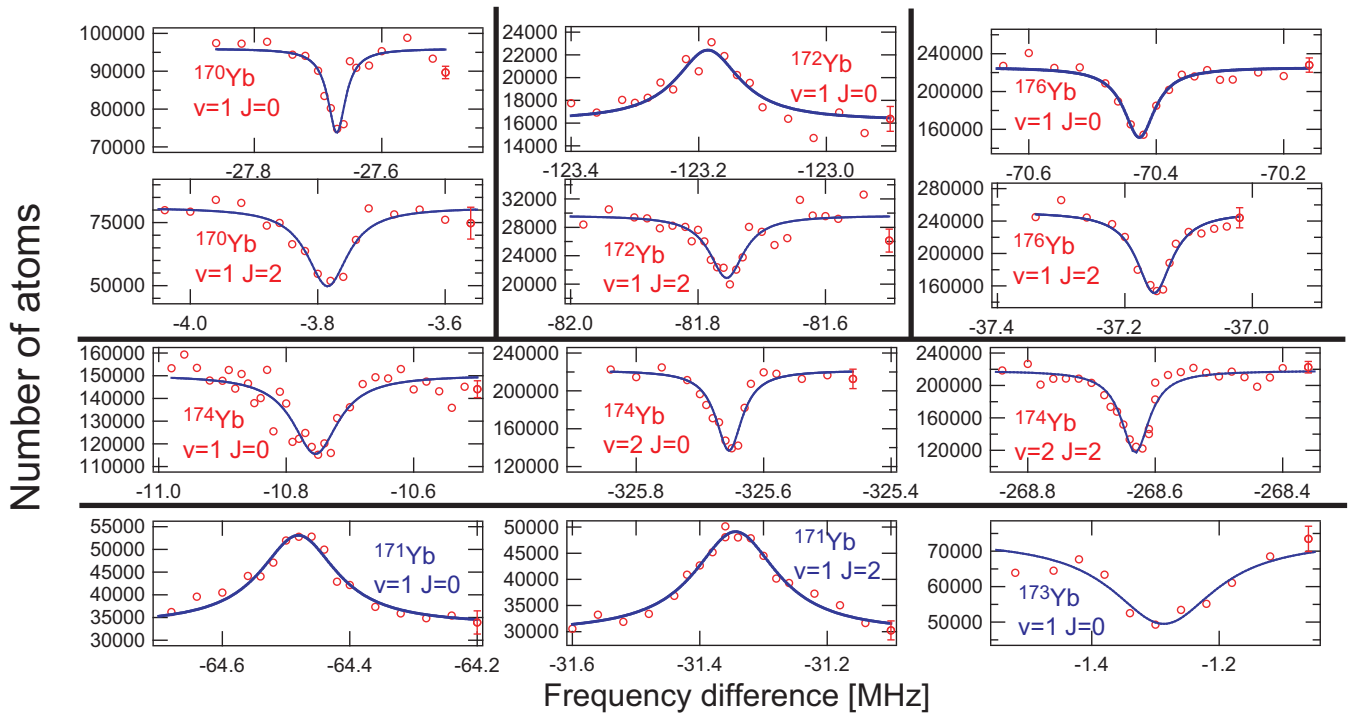}
\caption{(color online) Two-color PA spectra at about 1 $\mu$K. The horizontal and vertical axes are the frequency difference between the two lasers and the number of remained atoms, respectively.
The vertical error bars represent the fluctuations of the number of atoms.
The solid lines are fits of the Lorentz functions.
The Autler-Townes spectroscopy is applied to the $v=1, J=0,2$ states of $^{171}$Yb$_2$ and the $v=1, J=0$ state of $^{172}$Yb$_2$. while the Raman spectroscopy is applied to the others.}
\label{fig:spectra}
\end{center}
\end{figure*}

Ytterbium (Yb) is a rare-earth element with an electronic structure similar to that of the alkaline-earth atoms.
One of the unique features of Yb atoms is a rich variety of isotopes with five spinless bosons ($^{168}$Yb, $^{170}$Yb, $^{172}$Yb, $^{174}$Yb, $^{176}$Yb) and two fermions ($^{171}$Yb with the nuclear spin $I=1/2$ and $^{173}$Yb with $I=5/2$),
which enables us to study various mixtures of degenerate gases of Yb atoms.
In fact, recently, a BEC \cite{TakasuBEC} and a DFG \cite{FukuharaDFG} of Yb atoms have been achieved.
Another distinct feature of Yb atoms is the simple electronic ground state of $^1S_0$ symmetry. Therefore, the ground molecular state of Yb has only one potential of $^1\Sigma _g$ molecular symmetry with no electronic orbital and spin angular momenta.
This is in contrast to the case of an alkali dimer which has spin-singlet $^1 \Sigma _g$ and spin-triplet $^3 \Sigma _g$ ground states with a complicated hyperfine structure.
These two unique features of the Yb system, that is, the existence of rich variety of isotopes and one simple isotope-independent molecular potential,
offers the possibility to systematically check the scattering length theory with an unprecedented precision.
If we can be confident in the theory, then it would enable us to determine the scattering lengths of all possible isotope pairs which are not measured experimentally.
So far, to determine the $s$-wave scattering length of $^{174}$Yb, one-color PA spectroscopy was performed \cite{Enomoto}, which gave the result of $5.53(11)$ nm.
From the cross-dimensional rethermalization technique, the absolute value of the scattering length for $^{173}$Yb was estimated as
$6(2)$ nm \cite{FukuharaDFG}, assuming that the spin was completely unpolarized.
However, these two values are not enough for a rigorous test of the theory.

In this paper, we report an accurate determination of the $s$-wave scattering lengths for all Yb isotopes including those for different isotope pairs.
By using two-color PA spectroscopy with the intercombination transition $^1S_0-^{3\!\!}P_1$, we successfully determined the binding energy $E_b$ of twelve bound states near the dissociation limit of four homonuclear dimers comprised of bosonic atoms
 ($^{170}$Yb$_{2}$, $^{172}$Yb$_{2}$, $^{174}$Yb$_{2}$, and $^{176}$Yb$_{2}$) and two dimers comprised of fermionic atoms ($^{171}$Yb$_{2}$ and $^{173}$Yb$_{2}$).
The spectroscopically measured binding energies are in excellent agreement with theoretical calculations based on a simple model potential that was fit to the data.
The calculated $s$-wave scattering lengths for the six isotopes obey the mass-scaling law with very good precision.
Moreover, this excellent agreement allows us to accurately determine the scattering lengths of all twenty-eight different isotopic combinations.
In addition, we can reveal scattering properties of other partial waves such as $p$- and $d$-wave scatterings and energy dependence of the elastic cross sections.
These results are an important foundation for future research,
such as the efficiency of evaporative cooling, stability of quantum gases and their mixtures, and the clock shift \cite{Katori}.

\section{EXPERIMENT}

The experimental setup was almost the same as our previous experiment of one-color PA spectroscopy~\cite{Tojo}.
All the experiments were performed at about 1 $\mu$K, where only the $s$-wave scattering is possible.
Atoms were first collected in a magneto-optical trap (MOT) with the intercombination transition $^1S_0-^{3\!\!}P_1$ at $556$ nm.
The linewidth and saturation intensity of the transition were $182$ kHz and $0.14$ mW/cm$^2$, respectively.
The laser beam for the MOT was generated by a dye laser whose linewidth was narrowed to less than $100$ kHz. The dye laser was stabilized by an ultralow expansion cavity,
whose frequency drift was typically less than $20$ Hz/s.
The number, density and temperature of atoms in the MOT were about $2 \times 10^7$, $10^{11} $ cm$^{-3}$ and 40 $\mu $K, respectively.
Then the atoms were transferred into a crossed far-off resonant trap (FORT).
The number, density and temperature of atoms in the FORT were about $2 \times 10^6$, $10^{13} $ cm$^{-3}$, and 100 $\mu $K, respectively.
To reach lower temperatures, evaporative cooling was carried out by gradually decreasing the potential depth of the horizontal FORT beam to several tens of $\mu$K in several seconds.
Evaporative cooling worked rather well for the bosonic isotopes, $^{170}$Yb, $^{172}$Yb and $^{174}$Yb and for the fermionic isotope $^{173}$Yb.
Typically $1 \times 10^5$ atoms at the temperature of about 1 $\mu $K finally remained in the trap and
the density was between $10^{13} $ cm$^{-3}$ and $10^{14} $ cm$^{-3}$,
although optimized evaporation ramps and efficiency for each isotope were different.
In particular, we observed rapid atom decay for $^{172}$Yb due to three-body recombination.
The steeper evaporation ramp was needed for $^{172}$Yb to obtain enough number of atoms.
It is also noted that an unpolarized sample of $^{173}$Yb enabled us to perform efficient evaporative cooling
even at a low temperature via elastic collisions between atoms with
different spins.
For the bosonic $^{176}$Yb and fermionic $^{171}$Yb isotopes, however,
the evaporative cooling did not work well.
In order to cool $^{171}$Yb and $^{176}$Yb, we performed sympathetic cooling with bosonic $^{174}$Yb~\cite{Honda, Fukuhara}.
Bichromatic MOT beams for simultaneous trapping of two isotopes in the MOT were generated by an electro-optic modulator (EOM),
of which the modulation frequency corresponds to the isotope shift.
Typically $6 \times 10^4$ atoms for $^{171}$Yb and  $2 \times 10^5$ atoms for $^{176}$Yb at the temperature of about $1$ $\mu$K finally remained in the trap.
The bosonic isotope of $^{168}$Yb was hard to collect in the MOT due to its small natural abundance of 0.13 percent, although there are no fundamental difficulties for $^{168}$Yb in principle.

After the evaporative cooling, the two lasers, $L_1$ for the free-bound transition
and $L_2$ for the bound-bound transition, were simultaneously applied to the atoms in the trap for about 30 ms. These beams were focused to a $100$ $\mu$m diameter.
The schematic setup for the PA lasers is represented in Fig.~\ref{fig:PAlaser}.
The two laser beams were prepared by splitting one laser beam for the MOT
and therefore had the same frequency linewidth and stability as the MOT beam.
The relative frequency was controlled by acousto-optic modulators (AOMs).
The two laser beams were coupled in the same optical fiber.
A polarizing beam splitter (PBS) and a half-wave ($\lambda /2$) plate after the optical fiber were inserted to fix the polarization of both lasers in the same direction.
Finally, the PA laser beams were aligned to pass through the atoms in the FORT by using a CCD camera for absorption imaging.
The detuning of the PA laser with respect to the atomic resonance $^1S_0-^{3\!\!}P_1$ was easily checked by observing the frequency at which the atoms in the MOT disappeared.
The power of the PA laser was monitored by a photodiode.
For the Raman spectroscopy, the frequency $f_1$ of $L_1$ was fixed with a certain detuning $\Delta $ from a particular PA resonance, and the frequency $f_2$ of $L_2$ was scanned to search for the bound states of the ground state.
For the Autler-Townes spectroscopy, $f_1$ was fixed to a particular PA resonance, and $f_2$ was scanned.
Table~\ref{tab:excited} shows the excited state rovibrational levels used for the free-bound transition, where the resonance position is given in frequency detuning from the atomic  $^1S_0-^{3\!\!}P_1$ transition.
Our setup for the PA laser beams could provide enough power for detuning smaller than $1$ GHz.
To obtain enough power for detuning larger than $1$ GHz, the sideband of the EOM was used for the MOT while the carrier is used as the PA lasers.
We observed the two-color PA signals by measuring the number of surviving atoms with the absorption imaging method with the $^1S_0-^{1\!\!}P_1$ transition.

\begin{table}
\begin{center}
\caption{Excited state rovibrational Yb dimer levels used for the free-bound PA transition, where
$v_e$ and $J_e$ are the respective vibrational and rotational quantum numbers; $v_e$ is numbered from the dissociation limit. 
The position of the excited state level is given as detuning in MHz from the frequency of the atomic $^1S_0-^{3\!\!}P_1$ transition. For the fermionic isotopes the hyperfine level with the largest total atomic
angular momentum for the $^{3\!}P_1$ state defines the dissociation limit.}
\label{tab:excited}
\begin{tabularx}{80mm}{XXXc}
\hline \hline
       isotope&  $v_e$ & $J_e$ & position (MHz) \\ \hline
$^{170}$Yb  & 16 & 1 & 618      \\
$^{171}$Yb  & $a$ & $a$ & 904      \\
$^{172}$Yb  & 17 & 1 & 789   \\
$^{173}$Yb  &   &   & 791     \\
$^{174}$Yb  &18 & 1 & 993   \\
                  & 19 & 1 & 1972      \\
$^{176}$Yb   & 17 & 1 & 868      \\
 \hline \hline
$^a$Reference\cite{Enomoto2}.
\end{tabularx}
\end{center}
\end{table}

\section{EXPERIMENTAL RESULTS}

All the two-color PA spectra observed are shown in Fig.~\ref{fig:spectra}, where $v$ is the vibrational quantum number counting down from the dissociation limit and $J$ is the rotational quantum number.
The observed spectral linewidth is typically several tens of kHz due to
the finite energy distribution at the $1$ $\mu$K temperature and the power broadening.
The selection rule for the parity determines the observable quantum number $J$ in the ground state $^1\Sigma _g ^+$ as $0, 2, 4,\cdot \cdot \cdot $ from the $s$-wave collision.
We observed all resonance positions with detuning less than $300$ MHz from the dissociation limit except for the $v = 2$ state for $^{173}$Yb; see Fig.~\ref{fig:sl} (b).

\begin{figure}
\begin{center}
\includegraphics[width=8cm,clip]{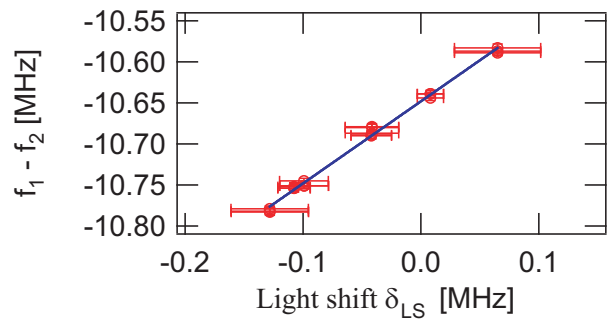}
\caption{(color online) Resonance positions $f_1-f_2$ of the $v=1, J=0$ state of $^{174}$Yb$_2$ as a function of $\delta_{LS}$.
The temperature shift is not compensated.
The vertical error bars represent the uncertainty estimates for the temperature
and the center frequency of the resonance.
The horizontal error bars include the fluctuations of the laser intensities.
The data are fitted with a linear function represented by a solid line.}
\label{fig:LightShift}
\end{center}
\end{figure}

\begin{figure}
\begin{center}
\includegraphics[width=8cm,clip]{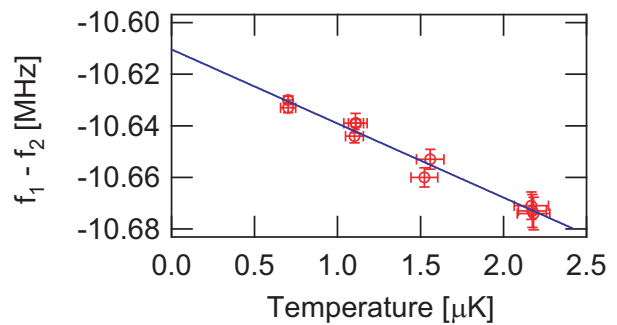}
\caption{(color online) Resonance positions $f_1-f_2$ of the $v=1, J=0$ state of $^{174}$Yb$_2$ as a function of the temperature.
The light shift is not compensated.
The vertical error bars represent the uncertainty estimates for the center frequency of the resonance.
The data are fitted with a linear function represented by a solid line.}
\label{fig:TempShift}
\end{center}
\end{figure}

The observed peak positions included the light shift.
We measured the peak positions with several laser intensities for $L_1$ and $L_2$ and different detunings $\Delta $, in order to compensate the light shift by interpolation or extrapolation.
We found that the bound-bound coupling is dominant for the light shift for the Raman spectroscopy.
Namely, the light shift is expressed as
\begin{eqnarray}
\delta_{LS} = \beta \left( \frac{I_1}{\Delta + f_1-f_2} + \frac{I_2}{\Delta} \right),
\label{LS}
\end{eqnarray}
where $\beta $ is a constant related to the Franck-Condon factor of the bound-bound transition,
$I_1$ and $I_2$ are the laser intensities of $L_1$ and $L_2$, respectively.
Typical values of the detuning $\Delta$ and laser intensity were about 2 MHz and $10-1000$ mW/cm$^2$.
The contributions of the other excited bound states were negligible,
since their resonance frequencies were typically more than $300$ MHz apart, which was much larger than $\Delta$ and $\Delta + f_1-f_2$.
Figure~\ref{fig:LightShift} shows the light shift $\delta_{LS}$ with several intensities for $L_1$ and $L_2$ and detuning $\Delta$.
The error bars include the uncertainty due to fluctuations of laser intensity and the uncertainty for the temperature of the atom clouds.
We note that the data are well fitted by a linear function.
It is also noted that the value of $\beta $ obtained from the fitting is consistent with our estimation of the Franck-Condon factor of the bound-bound transition, which also ensures the validity of our analysis.
The resonance position at $\delta_{LS} = 0$ gives the peak position without the light shift.
 For the Autler-Townes spectroscopy the light shift was basically caused by the laser $L_1$.
When the detuning $\Delta$ was not zero, however, the laser $L_2$ also contributed to the light shift.
The sense and magnitude of the detuning $\Delta$ could be estimated from the careful investigation of the spectral shape \cite{Bohn}.
Similarly, the peak positions with different laser intensities for $L_1$ and $L_2$ were measured and the light shift was compensated by extrapolation of these data.

\begin{table}
\caption{Measured and calculated binding energies $E_b$ for homonuclear isotopic pairs, where
$v$ and $J$ are the vibrational and rotational quantum numbers of the ground state dimer level and
$v$ is numbered from the dissociation limit.  R and AT respectively represent the Raman and Autler-Townes spectroscopic method of determining  $E_b$.
}
\label{tab:bind}
\begin{tabular}{ccccrrr}
\hline \hline
       isotope&  $v$ & $J$ & method & $E_b$ (MHz) & $E_b$ (MHz)  & Difference \\
       & & &  & experiment & theory &  (MHz)\\ \hline
$^{170}$Yb  & 1 & 0 & R  &  -27.661(23) &  -27.755 &  0.094 \\
            &   & 2 & R  &   -3.651(26) &   -3.683 &  0.032 \\
$^{171}$Yb  & 1 & 0 & AT &  -64.418(40) &  -64.548 &  0.130 \\
            &   & 2 & AT &  -31.302(50) &  -31.392 &  0.090 \\
$^{172}$Yb  & 1 & 0 & AT & -123.269(26) & -123.349 &  0.080 \\
            &   & 2 & R  &  -81.786(19) &  -81.879 &  0.093 \\
$^{173}$Yb  & 1 & 0 & R  &   -1.539(74) &   -1.613 &  0.074 \\
$^{174}$Yb  & 1 & 0 & R  &  -10.612(38) &  -10.642 &  0.030 \\
            & 1 & 0 & AT &  -10.606(17) &  -10.642 &  0.036 \\
            & 2 & 0 & R  & -325.607(18) & -325.607 &  0.000 \\
            & 2 & 2 & R  & -268.575(21) & -268.576 &  0.001 \\
$^{176}$Yb  & 1 & 0 & R  &  -70.404(11) &  -70.405 &  0.001 \\
            & 1 & 2 & R  &  -37.142(13) &  -37.118 & -0.024 \\
\hline \hline
\end{tabular}
\end{table}

\begin{table*}
\begin{center}
\caption{Calculated $s$-wave scattering lengths in nm for Yb isotopic combinations
}
\label{tab:scat}
\begin{tabularx}{160mm}{XXXXXXXX}
\hline \hline  \\
          &$^{168}$Yb&$^{170}$Yb&$^{171}$Yb&$^{172}$Yb&$^{173}$Yb&$^{174}$Yb&$^{176}$Yb\\
\hline
$^{168}$Yb&\bf 13.33(18)&6.19(8)&4.72(9)   &3.44(10)  &2.04(13)  &0.13(18)  &-19.0(1.6)\\
$^{170}$Yb&        &\bf 3.38(11)&1.93(13)  &-0.11(19) &-4.30(36) &-27.4(2.7)&11.08(12) \\
$^{171}$Yb&          &       &\bf -0.15(19)&-4.46(36) &-30.6(3.2)&22.7(7)   &7.49(8)   \\
$^{172}$Yb&          &          &      &\bf -31.7(3.4)&22.1(7)   &10.61(12) &5.62(8)   \\
$^{173}$Yb&          &          &          &       &\bf 10.55(11)&7.34(8)   &4.22(10)  \\
$^{174}$Yb&          &          &          &          &         &\bf 5.55(8)&2.88(12)  \\
$^{176}$Yb&          &          &          &          &          &       &\bf -1.28(23)\\
\end{tabularx}
\end{center}
\end{table*}

The observed peak positions also suffered from the temperature shift.
The temperature shift is assumed to be $a_T k_BT$, where $a_T$ is a constant, $k_B$ is the Boltzmann constant and $T$ is the temperature.
The factor $a_T$ was expected to be 3/2 \cite{He1},
which was checked experimentally for $v=1, J=0$ state of $^{174}$Yb$_2$ for the temperature range from 0.5 $\mu$K to 2 $\mu$K.
The resonance positions with different temperatures are shown in Fig.~\ref{fig:TempShift}.
The data were fitted with a linear function represented in a solid line.
The results are consistent with $\frac{3}{2} k_BT$ within 9 percent.
We also compensated the temperature shift for the other data using the $\frac{3}{2}k_BT$ dependence.
The final results of the experimentally determined binding energies
are listed in Table ~\ref{tab:bind}.

We also measured $E_b$ for the $v=1, J=0$ state of $^{174}$Yb$_2$
with the two-color PA using the strongly allowed
$^1S_0- ^{1\!\!}P_1$ transition.
The wavelength, linewidth, and saturation intensity of the
atomic transition were $399$ nm, $29$ MHz, and
$60$ mW/cm$^2$, respectively.
We used the $v_{e1} = 157$ level of the excited $^1\Sigma _u^+$ molecular state, where $v_{e1}$ is numbered from the dissociation
limit $^1S_0+ ^{1\!\!}P_1$ \cite{Enomoto}.
The resonance position of the level was $-172.3$ GHz from the dissociation limit.
Raman spectroscopy was employed to measure the peak position. % with different laser intensity for $L_1$ and $L_2$.
The light shift and temperature shift were also compensated.
The measured $E_b$ of $10.597(59)$ MHz  is in excellent agreement with the values in Table~\ref{tab:bind}, which includes the Autler-Townes measurement  with a precision of better than $20$ kHz.

\section{CALCULATION AND DISCUSSION}

The binding energies of the bound states as well as scattering lengths of all isotopic combinations are determined by the reduced mass and a single Born-Oppenheimer potential $V(r)$, 
as long as small mass-dependent adiabatic and non-adiabatic corrections to the potential are sufficiently small.  The key features of the potential that determine the positions of the last few bound states are the form of the long range potential and a phase associated with the short range potential.  
Consequently, we assume the following simple potential form to analyze the data:
\begin{equation}
V(r) = -\frac{C_6}{r^6}\left(1-\frac{\sigma^6}{r^6}\right)-\frac{C_8}{r^8}+B(r)J(J+1),
\label{LJ8}
\end{equation}
where $\sigma$ is a constant, $C_8$ is the van der Waals constant associated with the dipole-quadrupole interaction, and $B(r) = \hbar^2/(2\mu r^2)$ is due to molecular rotation.  
The first term in Eq.~(\ref{LJ8}) gives the Lennard-Jones form for the potential, for which the short range form can be changed by varying $\sigma$.   
The $C_8$ term is needed to improve the quality of the fit to the data.

\begin{figure}
\begin{center}
\includegraphics[width=9cm,clip]{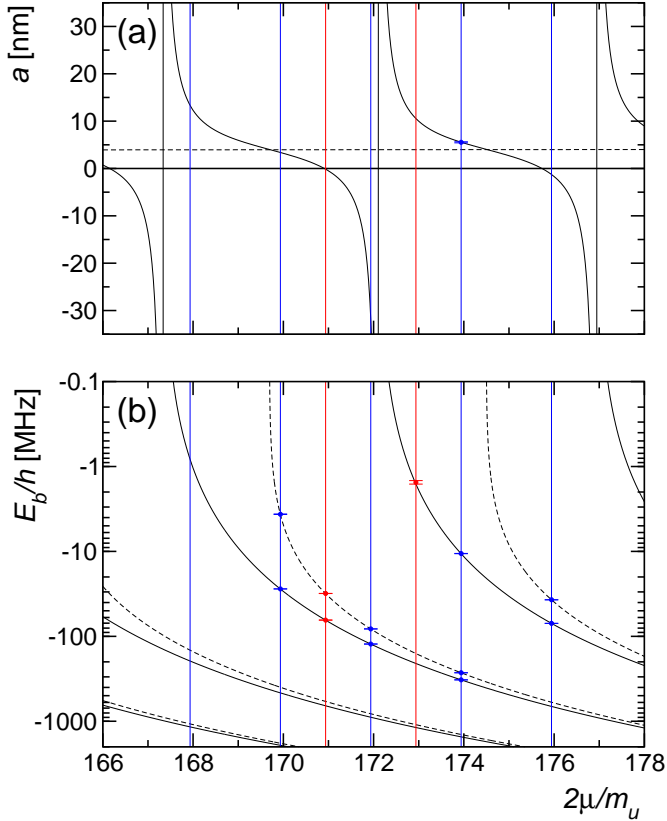}
\caption{(color online) Calculated scattering lengths (a) and binding energies $E_b$ (b) versus twice the reduced mass, 
$2\mu$ in the atomic mass unit, using the 3-parameter potential energy model in Eq.~(\ref{LJ8}) with the parameters given in the text.  
The vertical lines show the masses for the 7 like-atom pairs.  The solid and dashed lines in (b) show the $J=0$ and 2 eigenvalues respectively.  
The measured values with vertical error bars are shown for the levels for which they have been measured.  
The measured $^{174}$Yb scattering length of Ref.~\cite{Enomoto} is also shown in (a). The horizontal dashed line in (a) shows the van der Waals length $\bar{a}$. }
\label{fig:sl}
\end{center}
\end{figure}

By solving the Schr\"odinger equation numerically for the eigenvalues and comparing to the measured
binding energies, it is possible to determine an optimum set of potential parameters.  The $J=0$, $v=2$
level of $^{174}$Yb$_2$ and the $J=0$, $v=1$ level for $^{176}$Yb$_2$ were fit to determine $C_6$ and
obtain the right number of bound states $N_{174}$ for $^{174}$Yb$_2$. Then the $J=2$, $v=2$ level for
$^{174}$Yb$_2$ was added to the fit to determine $C_8$ and improve the determination of $C_6$.
The results are $C_6=1931.7\;E_{h}a_{0}^6$, $C_8=1.93\times 10^5\;E_{h}a_{0}^8$, $\sigma=9.0109362\;a_{0}$,
$N_{174}=72$, where $a_{0}\approx0.529\;{\rm nm}$, $E_{h}\approx4.36\times 10^{-18}\;{\rm J}$.
These parameters then determine without additional adjustment the binding energies of the other isotopic
combinations shown in Table~\ref{tab:bind} and Fig~\ref{fig:sl}(b).
Taking into account a lack of precise knowledge about the
short range part of the potential and the magnitude of possible retardation effects and neglect of
higher order dispersion energy terms, we estimate the uncertainties in $C_6$ and $C_8$ to be about
2 percent and 25 percent respectively, or $C_6=1932(30)\;E_{h}a_{0}^6$
and $C_{8}=1.9(5)\times 10^5\;E_{h}a_{0}^8$.
Adding a $-C_{10}/r^{10}$ van der Waals
term only very slightly improved the quality of the fit, and the optimum values of $C_6$ and $C_8$
remain within the above stated uncertainties.
The $C_6$ value is lower than a previous experimental determination of $2300(250) \;E_{h}a_{0}^6$ \cite{Enomoto} and recent theoretical predictions of $2291.6 \;E_{h}a_{0}^6$ \cite{Chu07},  
$2567.9 \;E_{h}a_{0}^6$ \cite{Buchachenko07}, and $2062 \;E_{h}a_{0}^6$ \cite{Zhang07}.
The short range form of the potential should be viewed as a pseudopotential having the right number
of bound states $N$ but not necessarily giving an accurate shape for the potential. The model well
depth $D_{e}/h=32469\;{\rm GHz}$ is significantly larger than {\it ab initio} values;
see Ref. \cite{Buchachenko07} and references therein.

The agreement shown in Table~\ref{tab:bind} with a precision less than
100 kHz between the experimentally determined $E_b$ and most of the
calculated $E_b$ values is quite impressive for such a simple model
potential.  The use of a single mass-independent potential with the
appropriate reduced mass is thus seen as an excellent approximation
for calculating the isotopic variation in binding energies.  The failure
to obtain a fit to the data within experimental error in all cases could
be indicative of the failure of mass scaling, although it may only be
due to the limitations of the simple form we assumed to represent the
potential over its whole range.   Adding a small mass-dependent
correction on the order of 1 GHz to the well depth of the potential
allows us to fit the binding energies for each isotope almost within
experimental uncertainties. We see indications that the potential is
deeper for heavier isotopes. Additional work is needed to
see if the Yb system can be used to make quantitative tests of the
accuracy and limitations of mass scaling.

When we added terms to account for relativistic retardation effects~\cite{Meath66}, the $\chi^2$ of the overall fit did not improve, although the energies of the least bound levels improved slightly.  
These relativistic terms take on the form $\alpha^2 W_4/r^4$ in the physically interesting region $r \ll \lambdabar$, 
where $\alpha$ is the fine structure constant and $\lambdabar \approx$ 1100 a$_0$ is a characteristic length associated with a mean electronic excitation energy.  
The lead term in the retarded van der Waals interactions switches to $1/r^7$ behavior in the very long range region $r \gg \lambdabar$.  
Since the outer turning point of the least bound level is still only around $150$ a$_0$, the $\alpha^2 W_4/r^4$ form is the appropriate form to use in looking for the magnitude of the effect.  
When we added a $C_4/r^4$ term to the potential, we find that we get a slightly better fit for the most weakly bound levels, but not a better overall fit, if we take $C_4=1.5\times 10^{-3}\;E_{h}a_{0}^3$, 
similar in magnitude to the value $0.6\times 10^{-3}\;E_{h}a_{0}^3$ estimated in the way proposed in Ref. \cite{Meath66} using the dipole polarizability and $C_6$ coefficient for Yb reported in Ref. \cite{Chu07}.   
Our reported model parameters and scattering lengths include the uncertainties associated with the lack of knowledge of the retardation corrections for this system.  
Consequently, there is a need for a theoretical evaluation of these corrections.

\begin{figure}
\begin{center}
\includegraphics[width=7.2cm,clip,,angle=270]{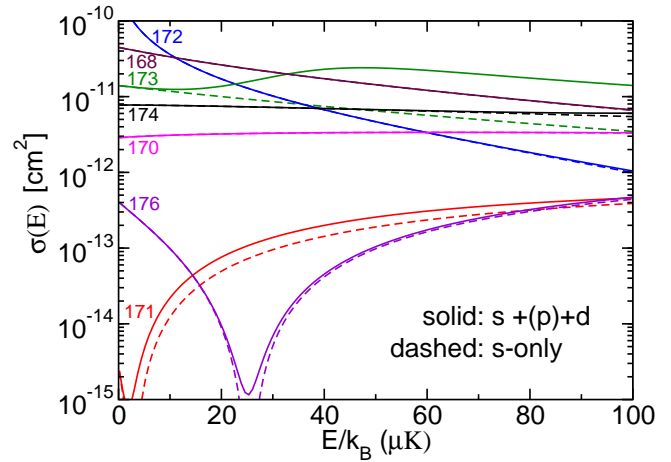}
\caption{(color online) Calculated cross section $\sigma(E)$ versus collision energy $E/k_B$ for two atoms of the same isotope. The label for each curve shows the isotopic mass number. Even mass number corresponds to identical bosons and odd mass number corresponds to two fermions with different spin projections.  The dashed lines show the $s$-wave contribution to the total cross section.  The solid lines show the contribution from $s$- and $d$- waves for the like boson cases, and the contribution from $s$-, $p$-, and $d$-waves for the fermion case with different spin components.  The $E \to 0$ cross sections are $8\pi a^2$ and $4 \pi a^2$ for the respective boson and fermion cases.  }
\label{fig:sigma}
\end{center}
\end{figure}

Given the ground state potential energy curve with the form of Eq.~(\ref{LJ8}) and the parameters given above, we can calculate the $s$-wave scattering lengths for all isotopic combinations by numerically solving the Schr{\"o}dinger equation with the appropriate reduced mass.  These are shown in Table~\ref{tab:scat} and Fig.~\ref{fig:sl}(a). The uncertainties reflect the uncertainties in the model parameters and the need to include retardation corrections to the potential.  The value for $^{174}$Yb is in excellent agreement with the value $5.53(11)$ reported in Ref.~\cite{Enomoto}.  The value for $^{173}$Yb is slightly larger than the value estimated from the thermalization experiments in Ref.~\cite{FukuharaDFG}.  However, the $s$-wave elastic cross section for $^{173}$Yb decreases by about 10 percent between $E=0$ and the 6 $\mu$K temperature of the experiment.  The experimental value could also be larger if the unknown distribution of spin populations differed from the assumption of uniformity.

Since the three-parameter model potential Eq. (\ref{LJ8}) is assumed to be common for all possible isotopic combinations, the simple analytical formula Eq. (\ref{massscaling}) also holds. The predictions of this formula are completely indistinguishable from those of the numerical calculation on the scale of Fig.~\ref{fig:sl} (a). The actual difference between scattering lengths calculated exactly from the model potential and those calculated from the analytical formula in Eq. (\ref{massscaling}) are below 0.03~nm for all like isotope cases except the one with the largest scattering length magnitude, $^{172}$Yb, for which the difference is 0.35~nm.  A similar statement applies to the mixed-isotope cases, for which most differences are below 0.03~nm.

As the reduced mass varies from 168/2 to 176/2, the scattering length varies through a complete cycle from $-\infty$ to $+\infty $.
In fact, we have a rich variety of the scattering lengths from large negative values for $^{172}$Yb -$^{172}$Yb, $^{171}$Yb -$^{173}$Yb, $^{170}$Yb -$^{174}$Yb, and $^{168}$Yb -$^{176}$Yb,
and almost zero for $^{171}$Yb -$^{171}$Yb, $^{170}$Yb -$^{172}$Yb, and $^{168}$Yb -$^{174}$Yb,
and to large positive for $^{172}$Yb -$^{173}$Yb and $^{171}$Yb -$^{174}$Yb.
Thus, the scattering length can be widely tuned by varying isotopic composition \cite{Takahashi}.

It should be noted that the observed behavior of evaporative cooling and sympathetic cooling are consistent with these scattering lengths.
The efficient evaporative cooling for $^{170}$Yb, $^{173}$Yb, and $^{174}$Yb is consistent with the large scattering lengths: $3.4$ nm for $^{170}$Yb, $11$ nm for $^{173}$Yb, and $5.6$ nm for $^{174}$Yb.
Inefficient evaporative cooling for $^{171}$Yb and $^{176}$Yb is also consistent with the small scattering lengths $-0.1$ nm for $^{171}$Yb and $-1.3$ nm for $^{176}$Yb.  On the other hand, we have successfully used sympathetic cooling to cool $^{171}$Yb and $^{176}$Yb with $^{174}$Yb, for which the respective mixed-isotope scattering lengths are 22 nm and 2.9 nm.
The extremely large value of $-32$ nm for $^{172}$Yb would explain very rapid atom decay observed for this system, which could be due to three-body recombination.

Using the three-parameter model potential, we can also calculate the collisional properties of other partial waves at non-zero collision energies. Figure~\ref{fig:sigma} shows the energy-dependent cross section for the collision of like isotopic species.  The cross section can be resonantly enhanced by a shape resonance, caused by the existence of quasibound rovibrational levels supported by the centrifugal barrier.  A $d$-wave shape resonance exists for $^{174}$Yb as already pointed out in Refs.~\cite{Tojo, Enomoto}. Our model predicts a broad peak in the collision cross section near $E/h=4.5$ MHz, or $E/k_B=220$ $\mu$K, which is off-scale in Fig.~\ref{fig:sigma}.  We found that a low energy $p$-wave shape resonance also exists for $^{173}$Yb, giving rise to a peak in the cross section near $E/k_B=48$ $\mu$K.

The $s$-wave contribution to the cross section becomes zero when the $s$-wave collisional phase shift is zero.   Such a zero as a function of energy results in a Ramsauer-Townsend minimum in the cross section.  This minimum is especially evident when the collision energy is so small that other partial waves make negligible contributions to the cross section,    This effect occurs at very low collision energy when the scattering length has a small negative value, as for  $^{171}$Yb and for $^{176}$Yb.  Figure~\ref{fig:sigma} shows the calculated cross section minima near 2 $\mu$K and 25 $\mu$K for these respective species.  This effect explains why evaporative cooling is found to be inefficient for these isotopes.

\section{CONCLUDING REMARKS}

In conclusion, we report the accurate determination of the binding energies of the twelve least bound states in the ground molecular potentials for six Yb isotopes
and the accurate determination of the $s$-wave scattering lengths for all possible combination of the isotopes based on a simple three-parameter model potential.
The model parameters are based on fitting a very limited set of the
experimental data, making use of only three binding energies: two
bound-states of $^{174}$Yb$_{2}$ and one bound-state of $^{176}$Yb$_{2}$.
Fitting this limited set of data allows us to reproduce the other nine binding energies with an
accuracy of about 100 kHz and to determine the 28 scattering lengths for the different
isotope combinations with an accuracy of about a few percent for most cases.
In addition, we can calculate the energy-dependent elastic scattering due to other partial waves such as $p$- and $d$- waves.
These results provide an important foundation for future research with Yb atoms on such topics as the efficiency of evaporative cooling, the stability of quantum gases and their mixtures, and clock shifts.

\begin{acknowledgments}
%We thank M. Kumakura for fruitful discussion.
We acknowledge
%Prof. Emeritus
S. Uetake, A. Yamaguchi, T. Fukuhara and H. Hara
for their experimental assistances.
We thank M. Portier and J. L\'eonard for helpful comments. 
This work was partially supported by
Grant-in-Aid for Scientific Research of JSPS
(18043013, 18204035), SCOPE-S,
and 21st Century COE "Center for
Diversity and Universality in Physics" from MEXT of Japan.
K. E. acknowledges support from JSPS.
The research is also part of the program of the National
Laboratory FAMO in Toru\'n, Poland.  PSJ acknowledges helpful comments from E. Tiemann and partial
support by the U. S. Office of Naval Research.
\end{acknowledgments}

\end{document}